\documentclass[11pt,twoside]{article}
\usepackage{asp2010}
\usepackage{natbib}
\usepackage{graphicx}

\resetcounters

\newcommand{\mic}{\mbox{$\,\mu$m}}

\newcommand{\fion}[2]{[{#1}\,{\sc {#2}}]}
\newcommand{\iso}{\mbox{\it Infrared Space Observatory}}

\newcommand{\spitzer}{\mbox{\it Spitzer}}

\begin{document}

\title{A {\it WISE} view of novae}
\author{A. Evans,$^1$ R. D. Gehrz,$^2$ C. E. Woodward,$^2$ and L. A. Helton$^3$
\affil{$^1$Astrophysics Group, Keele University, Keele, Staffordshire, ST5 5BG, UK}
\affil{$^2$Minnesota Institute for Astrophysics, School of Physics and Astronomy,\\
116 Church Street, S. E., University of Minnesota, Minneapolis, \\Minnesota 55455, USA}
\affil{$^3$SOFIA Science Center, USRA, NASA Ames Research Center, M.S. N232-11, Moffett Field, CA 94035, USA}}

\begin{abstract}
We present the result of trawling through the WISE archive for data on classical
and recurrent novae. The data show a variety of spectral energy distributions,
including stellar photospheres, dust and probable line emission. During the
mission WISE also detected some novae which erupted subsequent to the survey,
providing information about the progenitor sytems. 
\end{abstract}

\section{Introduction}

The Wide-Field Infrared Survey Explorer (WISE) was an all-sky mid-infrared
mission \citep{WISE}, that operated in wavebands having effective wavelengths
$\lambda=3.4, 4.6, 12$ and 22\mic. It commenced its survey on 2010 January 14,
completed its all-sky survey on 2010 July 17, and the mission terminated on 2011
February 1. In the course of its survey, WISE detected a number of classical and
recurrent novae and we have trawled through the WISE archive for Galactic novae.

Classical novae are well-known to go through a nebular (and in some cases,
coronal) phase when emission lines dominate the spectrum. Many of these lines,
which are known to be strong in mature novae \citep[see][and references
therein]{gehrz08,BASI,helton,GEW}, fall in the WISE bandpasses; for example, neon lines --
which are strongest in novae that originate on ONe white dwarfs -- affect WISE
Bands~1, 3 and~4. As discussed by \cite{martin}, infrared fine-structure lines
can remain strong for many decades after the nova outburst.

Also, as discussed by \cite{GEW} and \cite{helton_ct}, $\sim20$\% of novae (and
generally those that arise on CO white dwarfs) are dust formers. The
mineralogy of the dust is varied, with carbon, silicates and hydrocarbon
commonly seen. The dust temperature late in the evolution is typically several
100s of K, placing its emission in the WISE filters.

We have data-mined the WISE archive for classical and recurrent novae,
including mature and recent novae, and novae that have erupted since the WISE
mission. We present here the preliminary results of this trawl, and we
concentrate on a selection of objects with $>4\sigma$ detections in all four
WISE bands.

\section{Results}

\subsection{Recurrent novae with giant secondaries}

\begin{figure}[!ht]
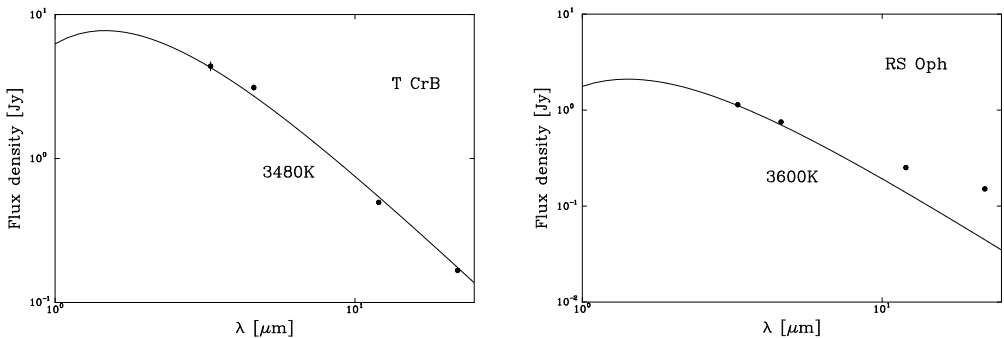

\plotfiddle{tcrb_1.eps}{2cm}{-90}{25}{25}{-200}{70}
\plotfiddle{rsoph_1.eps}{2cm}{-90}{25}{25}{0}{140}
\caption[]{Left: WISE SED of the recurrent nova T~CrB; note that there is
no evidence for an infrared excess. Right: The recurrent nova RS~Oph; note the
excess due to dust. In each case the temperature corresponds to the spectral
type of the giant component; curves are corresponding black bodies.
Error bars are smaller than the plotted points if they are not shown.
\label{RNE}}
\end{figure}

In Fig.~\ref{RNE} we show the WISE spectral energy distributions (SED) of T~CrB
and RS~Oph, two recurrent novae with giant secondaries; we have fitted the data
with black body curves having temperatures corresponding to the spectral types
of the secondaries. In the case of
T~CrB, the WISE data are consistent with the Rayleigh-Jeans tail of
the secondary, with no evidence for excess emission. However in the case of
RS~Oph there is clearly an excess in WISE bands~3 and~4, which we attribute to
the dust known to be present in the system
\citep{evans_rsoph,rushton,rushton_ct}. During its recurrent nova eruptions
\citep[see][and references therein]{RSOPH} there is evidence that the ejected
material interacts strongly with the giant secondary wind and the binary
environment \citep{mohamed}. There seems to be little evidence for circumstellar
material in the environment of T~CrB and it will be interesting to see (when its
long-awaited eruption occurs!) the extent to which the sequence of events that
occurred during the eruption of RS~Oph is replicated in T~CrB.

\subsection{Classical novae with evidence for line emission}

\begin{figure}[t]
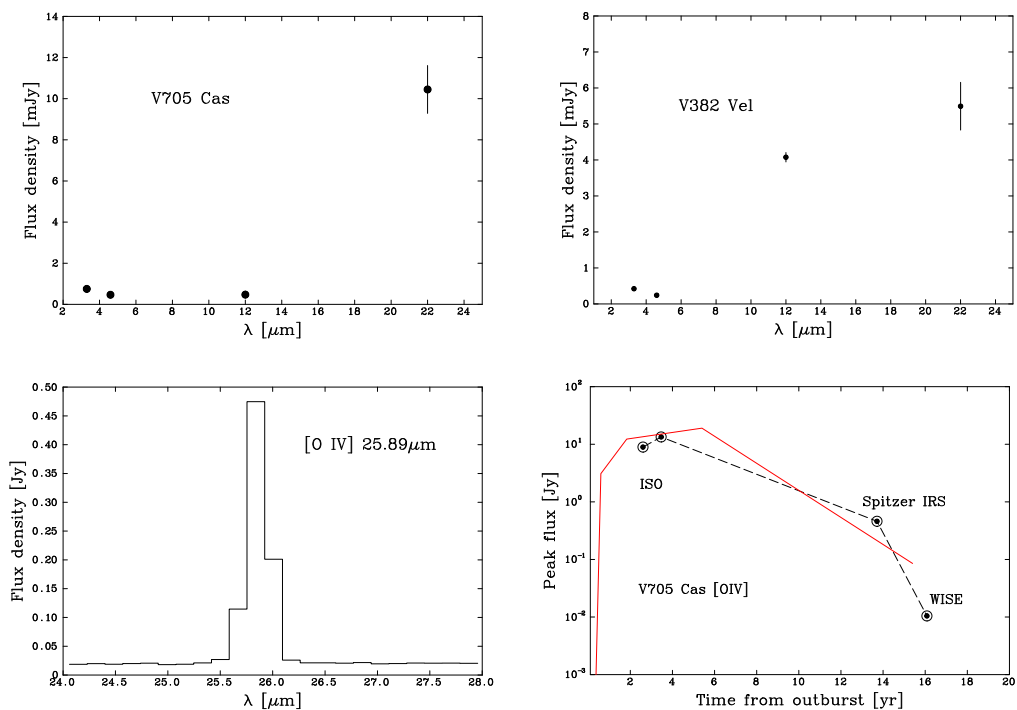


\plotfiddle{v705cas.eps}{2cm}{-90}{25}{25}{-200}{70}
\plotfiddle{v382vel.eps}{2cm}{-90}{25}{25}{0}{140}
\plotfiddle{cas_irs.eps}{2cm}{-90}{25}{25}{-200}{70}
\plotfiddle{cas_oiv.eps}{2cm}{-90}{25}{25}{0}{140}
\caption[]{Probable line emission in a CO nova (V705~Cas, top left) and an ONe
nova (V382~Vel, top right).
Error bars are smaller than the plotted points if they are not shown.
Bottom left: \spitzer\ IRS spectrum of V705~Cas obtained on 2007 August 29,
showing the \fion{O}{iv} 25.89\mic\ fine-structure line. 
Bottom right: decline in the peak flux in the \fion{O}{iv} line between 1996 and
2010 (dashed curve); solid curve is possible variation in the \fion{O}{iv} flux,
based on \citeauthor{martin}'s (\citeyear{martin}) analysis of DQ~Her and scaled
to the line flux in V705~Cas. \label{emission}} 
\end{figure}

In Fig.~\ref{emission} we show the WISE SEDs of V705~Cas (1993) and V382~Vel
(1999). The former was a dust-forming nova, the eruption occurring on the
surface of a CO white dwarf.
There is clearly a large excess in WISE Band~4 in the SED of V705~Cas,
which is very likely due to emission by the \fion{O}{iv} fine-structure line at
25.89\mic\ and which is ubiquitous in mature novae. This line was present in
V705~Cas in 1996, when it was observed with the \iso\ \citep{salama}, although
it was weak when it was observed by the {\it Spitzer Space Telescope}
\citep{werner,gehrz_spitzer}; the \spitzer\ IRS spectrum in the region of the
\fion{O}{iv} line is shown in Fig.~\ref{emission}.

Also shown in
Fig.~\ref{emission} is the decline in the peak flux in the \fion{O}{iv}
line in V705~Cas over the period 1996 (\iso) -- 2007 (WISE). In a comprehensive
study of the evolution of the line emission of DQ~Her (1934), \cite{martin}
estimated the evolution of the strengths of a number of recombination and
fine-structure lines for $\sim50$~years after eruption. With the caveat that
this analysis was for the specific case of DQ~Her, its stellar remnant and
ejecta abundances, the predicted evolution of the \fion{O}{iv} fins structure
line is in surprisingly good (qualitative) agreement with the evolution of the
\fion{O}{iv} line in V705~Cas.

V382~Vel, on the other hand, was an ONe nova \citep{woodward} which displayed
strong neon lines \citep{woodward,shore,helton}. The WISE SED for this object shows a
strong excess in Bands~3 and~4. The WISE data are superficially consistent with
the presence of strong fine-structure and coronal lines, specifically
\fion{Ne}{ii} 12.8\mic, \fion{Ne}{v} 14.3\mic\ and
\fion{Ne}{iii} 15.55\mic\ in Band~3, and \fion{Ne}{v} 24.32\mic, \fion{O}{iv}
25.89\mic\ in Band~4; all these lines were present in \spitzer\ spectra of
V382~Vel \citep{helton}. Further modelling is necessary to verify this and this is
in progress.

\subsection{Classical novae with evidence for dust emission}

\begin{figure}[!ht]
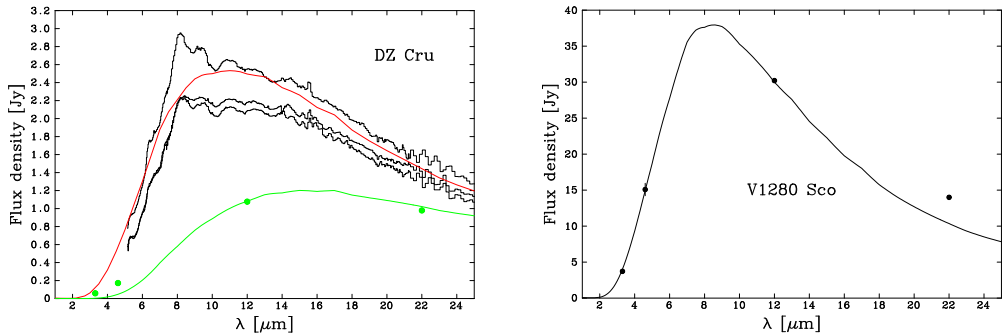

\plotfiddle{dzcru_sp.eps}{2cm}{-90}{25}{25}{-200}{70}
\plotfiddle{v1280sco_1.eps}{2cm}{-90}{25}{25}{0}{140}
\caption[]{Left: Evolution of the dust shell of DZ~Cru. Upper three spectra are from the {\it Spitzer Space Telescope}; the curve is a {\tt DUSTY} fit. WISE data are at the bottom of the diagram, with {\tt DUSTY} fit.
Right: Dust shell of V1280Sco, with {\tt DUSTY} fit.
Error bars are smaller than the plotted points if they are not shown.
See text for details. \label{dust}}
\end{figure}

In Fig.~\ref{dust} we show the WISE SEDs of DZ~Cru (2003) and V1280~Sco (2007),
both of which were prolific dust-producers, and both of which were observed by \spitzer.
Fig.~\ref{dust} includes the \spitzer\ data for DZ~Cru from
\cite{evans_dzcru}; these data were fitted with {\tt DUSTY} \citep{dusty} using
amorphous carbon at $\sim470$~K. The WISE data show that the dust emission is
shifting to longer wavelengths and declining in flux as the dust disperses.

We have attempted to fit the WISE data for DZ~Cru with {\tt DUSTY}. A constraint
is that the inner radius of the dust shell when DZ~Cru was observed by WISE (day~2351)
might be expected to be related to that in 2007 (day~874) when DZ~Cru was observed by
\spitzer, for example by the assumption of uniform outflow; the resulting dust
temperature is $\sim320$~K, although the fit is not entirely satisfactory (see Fig.~\ref{dust}).

The dust shell around V1280~Sco was observed at high spatial resolution by
\cite{chesneau}, who found that the dust resides in a bipolar configuration. The
WISE data suggest that the dust temperature at the inner edge of the shell was
$\sim500$~K in 2010 (see Fig.~\ref{dust}).

\subsection{Post-WISE novae}

\begin{figure}[t]
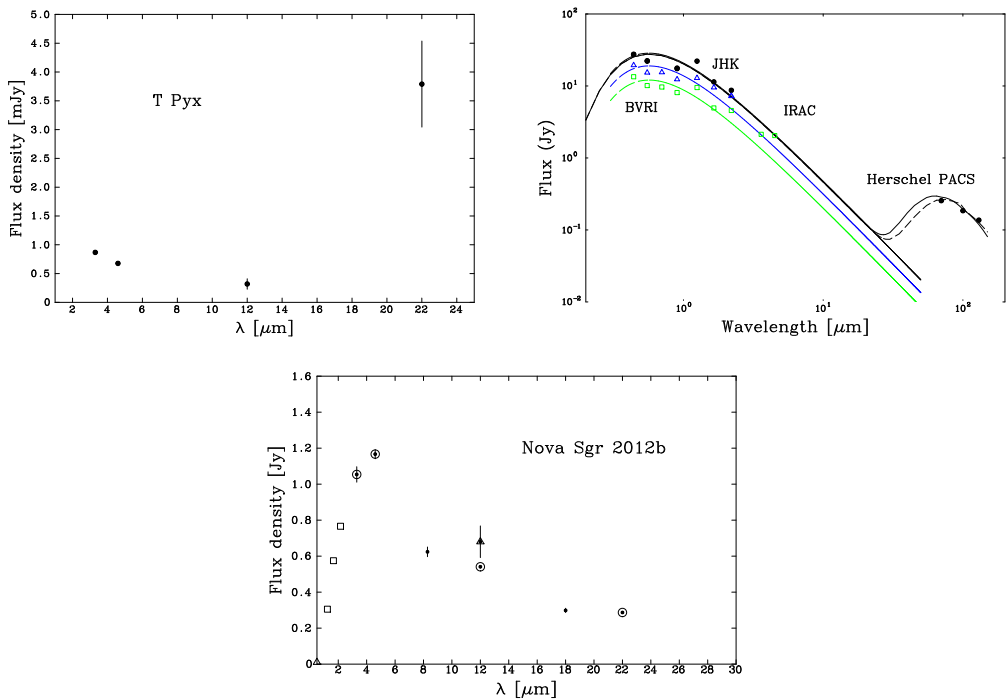

\plotfiddle{tpyx.eps}{2cm}{-90}{25}{25}{-200}{70}
\plotfiddle{tpyx_2011.eps}{2cm}{-90}{25}{25}{0}{140}
\plotfiddle{sgr2012b.eps}{4cm}{-90}{25}{25}{-100}{130}
\caption[]{Top left: The WISE SED of T~Pyx prior to its 2011 eruption; note the excess in
Band~4. Top right: The SED of T~Pyx during its 2011 eruption, including data from Mt. Abu, SMARTS, \spitzer\ IRAC and {\it Herschel} \citep[see][for details]{evans-tpyx}; note the excess longward of  $\sim50$\mic, and the difference in the flux levels before and during the eruption.
Bottom: WISE SED of the progenitor of Nova Sgr 2012b; WISE data have been supplemented by data from the {\it 2MASS} (open squares), {\it MSX} (small filled circles) and {\it IRAS} (triangle) catalogue.
Error bars are smaller than the plotted points if they are not shown.
See text for details.
\label{post-WISE}}

\end{figure}

In Fig.~\ref{post-WISE} we show the WISE SEDs of the recurrent T~Pyx and Nova
Sgr 2012b.

The SED of T~Pyx during the 2011 eruption is shown in Fig.~\ref{post-WISE} (top right). Data
from the {\it Herschel Space Observatory} \citep{herschel} show excess emission
longward of 50\mic, which \cite{evans-tpyx} attributed to interstellar dust that
had been swept up by a wind (either permanent, from the binary, or from nova
eruptions) from the T~Pyx binary. The WISE data for T~Pyx suggest an excess at
22\mic. Further work is needed to establish whether the same material is
responsible for the excess pre- and post-WISE.

The WISE SED of Nova Sgr 2012b is also shown in Fig.~\ref{post-WISE}, in which
the WISE data have been supplemented by data from {\it 2MASS, MSX} and {\it
IRAS}; we see strong emission which peaks at $\sim4$\mic. This is consistent
with optical spectroscopy of this nova, which reveals a symbiotic-like system
(F. M. Walter, private communication).

\section{Conclusions}

While the WISE survey does not have the broad wavelength coverage of the {\it
IRAS} survey it has vastly superior sensitivity and spatial resolution. It will
surely be as valuable a resource for investigating the circumstellar environment
of novae as was {\it IRAS} some 30~years ago \citep[e.g.][]{HG90}. Moreover it has the potential to
provide unprecedented information about nova {\em progenitors}.

Full details of this work will be published elsewhere.

\acknowledgements
RDG was supported by NASA and the US Air Force.
CEW acknowledges support from NASA.
We thank Alex d'Angelo for undertaking a preliminary trawl through the WISE
data.

\end{document}